\documentclass[prd,aps,twocolumn,floatfix]{revtex4}
\usepackage{hyperref}
\everymath={\displaystyle}
\usepackage{young}
\usepackage{rotating}
\usepackage{longtable}
\def\1{\mbox{l\hspace{-0.53em}1}}
\newcommand{\fr}{\frac}
\usepackage{multirow}
\newlength{\AccoHaut}
\newcommand{\Accolade}[3]{%
\setlength{\AccoHaut}{#1\baselineskip}
\setlength{\AccoHaut}{.5\AccoHaut}
\multirow{#1}{#2}{$\left.\rule{0pt}{\AccoHaut}\right\}$#3}}

\begin{document}
\title{A new look at the $[{\bf 70},1^-]$ baryon multiplet in the $1/N_c$ expansion}

\author{N. Matagne\footnote{Present address: Institut f\"ur Theoretische Physik, Universit\"at Giessen, D-35392 Giessen, Germany. \\ E-mail address: Nicolas.Matagne@theo.physik.uni-giessen.de}}

\author{Fl. Stancu\footnote{E-mail address: fstancu@ulg.ac.be
}}
\affiliation{University of Li\`ege, Institute of Physics B5, Sart Tilman,
B-4000 Li\`ege 1, Belgium}

\date{\today}

\begin{abstract}
So far, the  masses of excited states of mixed orbital symmetry and in 
particular those of nonstrange  $[{\bf 70},1^-]$  baryons   
derived in the $1/N_c$ expansion  were  
based on the separation of a system of $N_c$ quarks into
a symmetric core and an excited quark. Here we avoid this separation 
and show that an advantage of this new approach is to substantially 
reduce the number of linearly independent operators entering the 
mass formula. A novelty is that the isospin-isospin term
becomes as dominant in $\Delta$  as the spin-spin 
term in $N$ resonances.

\end{abstract}

\maketitle

 

\section{Introduction}
In 1974 't Hooft \cite{HOOFT} suggested a perturbative expansion
of QCD in terms of the parameter $1/N_c$  where $N_c$ is the number
of colors. This suggestion, together with 
the power counting rules of Witten \cite{WITTEN} has lead to
the   $1/N_c$ expansion method which allows to systematically  
analyze baryon properties. 
The current research status is described, for example, in
Ref. \cite{TRENTO}.
The success of the method stems from the discovery that the ground state baryons have 
an exact contracted SU(2$N_f$) symmetry 
when $N_c \rightarrow \infty $  \cite{Gervais:1983wq,DM},  $N_f$
being the number of flavors.
A considerable amount of work has been devoted to the ground state
baryons  \cite{DM,DJM94,DJM95,CGO94,Jenk1,JL95,DDJM96}.
For $N_c \rightarrow \infty $ the baryon masses are degenerate. 
For finite $N_c$ the mass splitting starts at order $1/N_c$.
Operator reduction rules simplify the $1/N_c$ expansion
\cite{DJM94,DJM95}.
It is customary to drop higher order corrections of order $1/N^2_c$.

It is thought that 't Hooft's suggestion \cite{HOOFT} would lead to
an $1/N_c$ expansion to hold in all QCD regimes. Accordingly,
the applicability of the approach to excited states is a subject of
current investigation. 

In the language of the constituent quark model
the excited states can be grouped into excitation bands 
with N = 1, 2, 3, etc. units of excitation energy. 
Among them, the N = 1 band,
or equivalently the $[{\bf 70},1^-]$ multiplet, 
has been most extensively studied,
either for $N_f$ = 2
\cite{CGKM,Goi97,PY1,PY2,CCGL,CaCa98,Pirjol:2003ye,Cohen:2003tb}
or for  $N_f$ = 3 \cite{SGS}. In the latter case,
first order corrections in SU(3) symmetry breaking 
were also included.  In either case, the conclusion was that the splitting starts 
at order $N^0_c$.

The N = 2 band contains  the $[{\bf 56'},0^+]$, $[{\bf 56},2^+]$,
$[{\bf 70},\ell^+]$ ($\ell$ = 0, 2) and  $[{\bf 20}, 1^+]$ multiplets.
There are no physical resonances associated to  $[{\bf 20}, 1^+]$.
The few studies related to  the N = 2 band concern the 
$[{\bf 56'},0^+]$ for $N_f$ = 2  \cite{CC00}, $[{\bf 56},2^+]$ 
for $N_f = 3$ \cite{GSS}  and  $[{\bf 70},\ell^+]$   for 
$N_f = 2$ \cite{MS2}, later extended to $N_f = 3$  \cite{Matagne:2006zf}.
The method has also been applied 
to highly excited nonstrange and strange 
baryons  belonging to  $[{\bf 56},4^+]$ \cite{MS1} which
is the lowest of the 17 multiplets of the N = 4 band \cite{SS94}.

The mass operator $M$ is defined as a linear combination of
independent operators $O_i$ 
\begin{equation}
\label{massoperator}
M = \sum_{i} c_i O_i,
\end{equation} 
where the coefficients $c_i$ are reduced matrix elements that
encode the QCD dynamics and are  
determined from a fit to the existing data.
Here we are concerned with nonstrange baryons only.
The building blocks of the operators $O_i$ are the 
SU(2$N_f$) generators $S_i$, $T_a$ and $G_{ia}$ and the 
SO(3) generators $\ell_i$. 
Their general form is
\begin{equation}\label{OLFS}
O_i = \frac{1}{N^{n-1}_c} O^{(k)}_{\ell} \cdot O^{(k)}_{SF},
\end{equation}
where  $O^{(k)}_{\ell}$ is a $k$-rank tensor in SO(3) and  $O^{(k)}_{SF}$
a $k$-rank tensor in SU(2)-spin, but invariant in SU($N_f$).
Thus $O_i$ are rotational invariant.
For the ground state one has $k = 0$. The excited
states also require  $k = 1$ and $k = 2$ terms.

The spin-flavor (SF) operators  $O^{(k)}_{SF}$ are combinations 
of SU(2$N_f$) generators, the lower 
index $i$ in the left hand side of (\ref{OLFS}) representing a specific 
combination.
Each $\it n$-body operator  is multiplied by an explicit  factor of 
$1/{N^{n-1}_c}$ resulting from the power counting rules.
Some compensating $N_c$ factors may arise in the matrix elements when 
$O_i$ contains a coherent operator such as $G^{ia}$ or $T^a$.

The excited states belonging to  $[{\bf 56},\ell]$ multiplets are rather
simple and can be studied by analogy with the ground state. In this case 
both the orbital 
and the spin-flavor parts of the wave function are symmetric.
Naturally, it turned out that the splitting starts at order $1/N_c$ \cite{GSS,MS1},
as for the ground state.

The states belonging to $[{\bf 70}, \ell]$ multiplets are apparently more 
difficult.  So far, the general practice 
was to decouple the baryon into an excited quark and a symmetric core.
This means that each generator of SU($2N_f$) must be written as a sum of two
terms, one acting on the excited quark and the other on the core. 
As a consequence, the number of linearly independent operators $O_i$ increases 
tremendously  and the number of coefficients $c_i$, 
to be determined, becomes much larger than the experimental data
available. For example, for the $[{\bf 70},1^-]$ multiplet with $N_f = 2$ 
one has 13 linearly independent
operators up to order $1/N_c$  included \cite{CCGL}, instead of 7 
(see below).
We recall that there are only 7 nonstrange 
resonances belonging to this band.
Consequently, selecting the most dominant operators is very difficult so that
one risks to make an arbitrary choice  \cite{CCGL}.

In this practice
the matrix elements of the excited quark are straightforward, as being
described by single-particle operators. The matrix elements of the core operators 
$S^i_c, T^a_c$ are also simple to calculate, while those of $G^{ia}_c$
are more involved.
Analytic formulas for the matrix elements of all SU(4)
generators  have been derived  in Ref. \cite{HP}. Every matrix element is factorized 
according to a generalized Wigner-Eckart theorem into a reduced
matrix element and an SU(4) Clebsch-Gordan coefficient. These matrix elements have 
been used in nuclear physics, which is governed by the SU(4) symmetry.
Recently we have extended the approach 
of Ref. \cite{HP}  to SU(6)  \cite{Matagne:2006xx} and obtained matrix elements of
all SU(6) generators between symmetric $[N_c]$ states.

Here we propose a method where no decoupling is necessary. All one needs 
to know are the matrix
elements of the SU(2$N_f$) generators between mixed symmetric
states   $[N_c-1,1]$. For SU(4) they were obtained by 
Hecht and Pang \cite{HP}. They can be easily applied to  
a system of $N_c$ nonstrange quarks. 
To our knowledge such matrix elements are yet 
unknown for  $N_f$ = 3. 

\section{The wave function}

We deal with a system of $N_c$ quarks having one unit of
orbital excitation. Then the orbital wave function must have a mixed 
symmetry $[N_c-1,1]$. Its spin-flavor part 
must have the same symmetry in order to obtain a totally symmetric
state in the orbital-spin-flavor space. 
The general form of such a wave function is \cite{book}
\begin{equation}
\label{EWF}
|[N_c] \rangle = {\frac{1}{\sqrt{d_{[N_c-1,1]}}}}
\sum_{Y} |[N_c-1,1] Y \rangle_{O}  |[N_c-1,1] Y \rangle_{FS}
\end{equation}    
where $d_{[N_c-1,1]} = N_c - 1$ is the dimension of the representation 
$[N_c-1,1]$ of the permutation group $S_{N_c}$ and $Y$ is a symbol for a
Young tableau (Yamanouchi symbol). 
The sum is performed over all possible standard Young tableaux. 
In each term the first basis vector represents the orbital  
space ($O$) and the second the spin-flavor space ($FS$).
In this sum there is only one $Y$ (the normal Young tableau) 
where the last particle is in 
the second row and $N_c - 2 $ terms where the last particle
is in the first row. All these terms were neglected in the 
procedure of decoupling the excited quark,
which implies that the permutation symmetry $S_{N_c}$
was broken, {\it i.e.}  the orbital-spin-flavor wave function was no more
symmetric, as it should be.  One can easily prove the above assertion by 
looking at the expression of the wave function, Eqs. (3.4)-(3.5) 
in the second paper of 
Ref.  \cite{CCGL}, for example. This definition contains
the coefficients $c_{\rho \eta}$ which are defined as
coefficients of an ``orthogonal rotation". 
In Ref.  \cite{MS2} we have shown that  $c_{\rho \eta}$
are some specific isoscalar factors of the permutation group $S_{N_c}$ 
\cite{book}.
These are factors of the Clebsch-Gordan coefficients,
 factorized as isoscalar factors times  Clebsch-Gordan 
coefficients of the group $S_{N_c-1}$.
In the case under concern the isoscalar factors incorporate 
the position of the $N_c$-th particle in a Young tableau. 
By identifying our expressions with those of Ref.  \cite{CCGL}
we found that they correspond to the term where the last particle is located
in the second row of the Young tableau of the representation
$[N_c-1,1]$. Thus the other $N_c - 2$ terms of the wave function,
with the $N_c$-th particle in the first row, are missing. 
In Appendix A we show explicitly which are the missing terms for 
$N_c = 3$ in the sectors $^28$, $^48$ and $^210$. 
In addition, as an example, the orbital basis vectors of 
configuration $s^4 p$, containing one unit of orbital excitation,
which span the invariant subspace of the irreducible
representation $[41]$ of $S_5$ are given in Appendix B.
The definition and the orthogonality properties
together with examples of isoscalar factors can be 
found in Ref. \cite{ISOSC}. 
In Sec. VI we discuss the validity of the approximate (asymmetric) wave 
function of Ref. \cite{CCGL}.

If there is no decoupling, there is no need to
specify $Y$, the matrix elements being identical for all $Y$'s,
due to Weyl's duality between a linear group and a symmetric group
in a given tensor space \footnote{see Ref. \cite{book}, Sec 4.5.}.
Then the explicit form of a  wave function of total angular momentum 
$\vec{J} = \vec{\ell} + \vec{S}$ and  isospin $I$ is
\begin{eqnarray}\label{WF}
\lefteqn{|\ell S I I_3;JJ_3 \rangle  = 
\sum_{m_\ell,S_3}
      \left(\begin{array}{cc|c}
	\ell    &    S   & J   \\
	m_\ell  &    S_3  & J_3 
      \end{array}\right)} \nonumber \\
& & \times 
|[N_c-1,1]\ell m_{\ell} \rangle
~|[N_c-1,1] S S_3 I I_3 \rangle,
\end{eqnarray}
each term containing an SU(2) Clebsch-Gordan (CG) coefficient, an 
orbital part 
$|[N_c-1,1]\ell m_{\ell} \rangle$ an a spin-flavor
part $|[N_c-1,1] S S_3 I I_3 \rangle$.

\section{SU(4) generators as tensor operators}

The SU(4) generators $S_i$, $T_a$  and $G_{ia}$, 
globally denoted by $E_{ia}$ \ \cite{HP},
are components of an irreducible tensor operator
which transform according to the adjoint representation $[211]$ of
dimension $\bf 15$ of SU(4).
We recall that the SU(4) algebra is
\begin{eqnarray}\label{ALGEBRASU4}
&[S_i,T_a] = 0,
~~~~~ [S_i,G_{ja}]  =  i \varepsilon_{ijk} G_{ka},\nonumber \\
&~~~~~ [T_a,G_{ib}]  =  i \varepsilon_{abc} G_{ic},\nonumber \\
&[S_i,S_j]  =  i \varepsilon_{ijk} S_k,
~~~~~ [T_a,T_b]  =  i \varepsilon_{abc} T_c,\nonumber \\
&[G_{ia},G_{jb}] = \fr{i}{4} \delta_{ij} \varepsilon_{abc} T_c
+\fr{i}{4} \delta_{ab}\varepsilon_{ijk}S_k.
\end{eqnarray}
As one can see, the tensor operators $E_{ia}$ 
are of three types:
$E_i$ ($i$ = 1,2,3) which form the subalgebra of SU(2)-spin, 
$E_a$ ($a$ = 1,2,3) which form the subalgebra of SU(2)-isospin
and  $E_{ia}$ which act both in the spin and the isospin spaces. They
are related to $S_i$, $T_a$ and $G_{ia}$ $(i=1,2,3;\ a=1,2,3)$ by
\begin{equation} \label{normes2}
E_i =\frac{S_i}{\sqrt{2}};~~~ E_a = \frac{T_a}{\sqrt{2}}; ~~~E_{ia} = \sqrt{2} G_{ia}.
\end{equation}

The matrix elements of every $E_{ia}$
between states belonging to the representation $[N_c-1,1]$ 
can be expressed as a generalized Wigner-Eckart theorem
which reads\ \cite{HP}
\begin{eqnarray}\label{GENsu4}
\lefteqn{\langle [N_c-1,1] I' I'_3 S' S'_3 | E_{ia} |
[N_c-1,1] I I_3 S S_3 \rangle  =}  \nonumber \\ & &  \sqrt{C^{[N_c-1,1]}(\mathrm{SU(4)})}   
 \left(\begin{array}{cc|c}
	S   &    S^i   & S'   \\
	S_3  &   S^i_3   & S'_3
  \end{array}\right)
   \left(\begin{array}{cc|c}
	I   &   I^a   & I'   \\
	I_3 &   I^a_3   & I'_3
   \end{array}\right)\nonumber \\& &  \times 
    \left(\begin{array}{cc||c}
	[N_c-1,1]    &  [211]   & [N_c-1,1]   \\
	S I  &   S^i I^a   &   S' I'
      \end{array}\right)_{\rho=1},
   \end{eqnarray}
where $C^{[N_c-1,1]}(\mathrm{SU(4)})=N_c(3N_c+4)/8$ is the eigenvalue of the 
SU(4) Casimir operator for the representation $[N_c-1,1]$. 
The other three factors are:  
an  SU(2)-spin CG
coefficient, an SU(2)-isospin CG coefficient and an isoscalar factor of SU(4). 
Note that the isoscalar factor carries a lower index $\rho$ = 1.
In general, this index is necessary to distinguish 
between irreducible representations, whenever the multiplicity  
in the inner product $[N_c-1,1] \times [211] \rightarrow [N_c-1,1]$ is larger
than one. In that case, the matrix elements 
of the SU(4) generators in a fixed irreducible representation $[f]$ 
are defined such as the reduced matrix elements take the following values
\cite{HP}
{\small
\begin{equation}
\langle [f] || \mathrm{E} || [f] \rangle = 
\left\{
\begin{array}{ll}
\sqrt{C^{[N_c-1,1]}(\mathrm{SU(4)})} & \rm{for} \hspace{3mm} \rho = 1 \\
0 \hspace{3mm} & \rm{for} \hspace{3mm} \rho \neq 1 \\
\end{array}\right. .
\end{equation}}

\noindent Thus the knowledge of the matrix elements of SU(4) generators amounts to the
knowledge of the corresponding SU(4) isoscalar factors. In Ref. \cite{HP} 
a variety of isoscalar factors were obtained. We need those 
for $[f] = [N_c-1,1]$. 
They are reproduced  in Table \ref{ISOSC} in terms of our notation
and  typographical errors corrected.
They contain the phase factor introduced in Eq. (35) of Ref.  \cite{HP}. 
As compared to the symmetric  $[N_c]$ representation, where  $I = S$ always, 
here one has  $ I = S $ (13 cases) 
but also $I \neq S$ (10 cases). Some of the  properties 
of these isoscalar factors are given in Appendix C.

 One can easily identify the matrix elements associated 
 to the generators of SU(4). One has $ S_2 I_2 $ = 10 for $S_i$,
 $S_2 I_2 $ = 01  for $T_{a}$ and  $ S_2 I_2 $ = 11 for $G_{ia}$,
 where 1 or 0 is the rank of the SU(2)-spin or SU(2)-isospin tensor
 contained in the generator. 
The generalized Wigner-Eckart theorem (\ref{GENsu4}) is used to calculate the 
matrix elements of $O_i$  needed for the mass operator, as described
below.  
 

\begin{table*}
\begin{center}
{\scriptsize
\caption{Isoscalar factors of SU(4) for $[N_c-1,1] \times [211] 
\rightarrow [N_c-1,1]$ defined by Eq. (\ref{GENsu4}).}
\label{ISOSC}
\begin{tabular}{ll|c|c|cp{7cm}}
\hline
\hline
\vspace{0cm} 
\hspace{3cm}&    &\hspace{1.5cm}    &    & \hspace{1cm} \mbox{}& \\
$S_1$&$I_1$ \hspace{1.5cm} & \hspace{0cm}$S_2I_2$ \hspace{1.5cm}& \hspace{1.5cm}
$SI$ \hspace{1.5cm} & &\hspace{0cm}$\left(\begin{array}{cc||c}                                         [N_c-1,1]  &  [211]  &  [N_c-1,1] \\
                           S_1I_1 & S_2I_2 & SI
                                      \end{array}\right)_{\rho=1}$  \\
\vspace{0cm} &    &    &   &  & \\
\hline
\vspace{0cm} &    &    &    & & \\
$S+1$&$S+1$ & $11$ & $SS$ & & $\sqrt{\frac{S(S+2)(2S+3)(N_c-2-2S)(N_c+2+2S)}{(2S+1)(S+1)^2N_c(3N_c+4)}}$ \\ 
     &    &       &      && \\
$S+1$&$S$ & $11$  & $SS$ && \raisebox{-0.05cm}[0cm][0cm]{\hspace{-3cm}\Accolade{3}{8cm}{\hspace{2.52cm}$-\frac{1}{S+1}\sqrt{\frac{(2S+3)(N_c+2+2S)}{(2S+1)(3N_c+4)}}$}} \\ 
     &    &       &      && \\
$S$&$S+1$ & $11$  & $SS$ && \\
  \vspace{-0.2cm}   &    &       &      && \\
$S$&$S$   & $11$  & $SS$ & &$-\frac{N_c-(N_c+2)S(S+1)}{S(S+1)\sqrt{N_c(3N_c+4)}}$ \\
  \vspace{-0.2cm}    &    &       &      && \\
$S$&$S-1$ & $11$  & $SS$ & & \raisebox{-0.05cm}[0cm][0cm]{\hspace{-3cm}\Accolade{3}{8cm}{\hspace{2.52cm}$ -\frac{1}{S}\sqrt{\frac{(2S-1)(N_c-2S)}{(2S+1)(3N_c+4)}}$}}\\
     &    &       &      && \\
$S-1$&$S$ & $11$  & $SS$ & & \\
   \vspace{-0.2cm}   &    &       &      && \\
$S-1$&$S-1$ & $11$ & $SS$ && $\frac{1}{S}\sqrt{\frac{(S-1)(S+1)(2S-1)(N_c+2S)(N_c-2S)}{(2S+1)N_c(3N_c+4)}}$ \\
   \vspace{-0.2cm}   &    &       &      && \\
$S+1$&$S$   & $10$ & $SS$ && \raisebox{-0.08cm}[0cm][0cm]{\hspace{-3cm}\Accolade{3}{8cm}{\hspace{2.52cm}$0$}}\\
     &    &       &      && \\
$S$&$S+1$   & $01$ & $SS$ && \\
 \vspace{-0.2cm}    &    &       &      && \\
$S-1$&$S$   & $10$ & $SS$ & & \raisebox{-0.06cm}[0cm][0cm]{\hspace{-3cm}\Accolade{3}{8cm}{\hspace{2.52cm}$0$}}\\
    &    &       &      && \\
$S$&$S-1$   & $01$ & $SS$ &&  \\
  \vspace{-0.2cm}  &    &       &      && \\
$S$&$S$     & $10$ & $SS$ & & \raisebox{-0.05cm}[0cm][0cm]{\hspace{-3cm}\Accolade{3}{8cm}{\hspace{2.52cm}$\sqrt{\frac{4S(S+1)}{N_c(3N_c+4)}}$}} \\
    &    &       &      && \\
$S$&$S$     & $01$ 
& $SS$ & & \\
 \vspace{-0.2cm}     &    &       &      && \\
$S+1$&$S$   & $11$ & $SS-1$ && $\sqrt{\frac{(2S+3)(N_c+2+2S)(N_c-2S)}{(2S+1)N_c(3N_c+4)}}$ \\
 \vspace{-0.2cm}   &    &       &      && \\
$S$&$S$     & $11$ & $SS-1$ && $\frac{1}{S}\sqrt{\frac{N_c-2S}{3N_c+4}}$ \\
 \vspace{-0.2cm}   &    &       &      && \\
$S$&$S-1$   & $11$ & $SS-1$ & &$\frac{1}{S}\sqrt{\frac{(S-1)(S+1)N_c}{3N_c+4}}$ \\
 \vspace{-0.2cm}   &    &       &      && \\
$S-1$&$S$   & $11$ & $SS-1$ && $-\frac{N_c+4S^2}{S \sqrt{(2S-1)(2S+1)N_c(3N_c+4)}}$ \\
 \vspace{-0.2cm}   &    &       &      && \\
$S-1$&$S-1$ & $11$ & $SS-1$ && $\frac{1}{S}\sqrt{\frac{N_c+2S}{3N_c+4}}$ \\
 \vspace{-0.2cm}   &    &       &      && \\
$S-1$&$S-2$ & $11$ & $SS-1$ && $\sqrt{\frac{(2S-3)(N_c+2-2S)(N_c+2S)}{(2S-1)N_c(3N_c+4)}}$ \\
 \vspace{-0.2cm}   &    &       &      && \\
$S$&$S-1$   & $10$ & $SS-1$ & &$\sqrt{\frac{4S(S+1)}{N_c(3N_c+4)}}$ \\
 \vspace{-0.2cm}   &    &       &      && \\
$S-1$&$S-1$ & $10$ & $SS-1$ & &0 \\
 \vspace{-0.2cm}   &    &       &      && \\
$S$&$S$     & $10$ & $SS-1$ & &0 \\
 \vspace{-0.2cm}   &    &       &      && \\
$S$&$S-1$   & $01$ & $SS-1$ & &$\sqrt{\frac{4(S-1)S}{N_c(3N_c+4)}}$ \\   &    &       &      && \\
\hline
\hline
\end{tabular}}\end{center}
\end{table*}


\section{The mass operator}

As specified in the introduction, here we are concerned with 
nonstrange baryons only.
Table  \ref{operators} contains the seven independent 
operators up to order $1/N_c$
appearing in the mass operator  Eq. (\ref{massoperator}).
As already mentioned, the building  blocks of  $O_i$ 
are  $S^i$, $T^a$, $G^{ia}$ and $\ell^i$. 
We also need the rank $k = 2$ tensor operator 
\begin{equation}\label{TENSOR}
\ell^{(2)ij} = \frac{1}{2}\left\{\ell^i,\ell^j\right\}-\frac{1}{3}
\delta_{i,-j}\vec{\ell}\cdot\vec{\ell},
\end{equation}
which, like $\ell^i$, acts on the orbital wave function $|\ell m_{\ell} \rangle$  
of the whole system of $N_c$ quarks (see  Ref. \cite{MS2} for the normalization 
of $\ell^{(2)ij}$). 

\begin{table*}[h!]
\caption{List of operators and the coefficients resulting from 
numerical fits. The values of $c_i$ are indicated under the headings Fit n,
in each case.}
\label{operators}{\scriptsize
\renewcommand{\arraystretch}{2} 
\begin{tabular}{lrrrrrr}
\hline
\hline
Operator \hspace{2cm} &\hspace{0.0cm} Fit 1 (MeV) & \hspace{0.5cm} Fit 2 (MeV) & \hspace{0.5cm}Fit 3 (Mev) &\hspace{0.5cm} Fit 4 (MeV) &\hspace{0.5cm} Fit 5 (MeV)&\hspace{0.5cm} Fit 6 (MeV) \\
\hline

$O_1 = N_c \ \1 $                            & $481 \pm5$  & $482\pm5$ &  $484\pm4$ &  $484\pm4$ & $498\pm3$ & $495\pm3$ \\
$O_2 = \ell^i s^i$                	     & $-31 \pm26$ & $-20\pm23$ & $-12\pm20$ & $3\pm15$ & $38\pm34$ & $-30\pm 25$ \\
$O_3 = \frac{1}{N_c}S^iS^i$                  & $161\pm 16$ & $149\pm11$ & $163\pm16$ & $150\pm11$ & $156\pm16$ & \\
$O_4 = \frac{1}{N_c}T^aT^a$                  & $169\pm36$  & $170\pm36$ & $141\pm27$ & $139\pm27$ & \\
$O_5 = \frac{15}{N_c}\ell^{(2)ij}G^{ia}G^{ja}$    & $-29\pm31$&            & $-34\pm30$&        & $-34\pm31$ & $-32 \pm 29$\\
$O_6 = \frac{3}{N_c}\ell^iT^aG^{ia}$            & $32\pm26$ & $35\pm26$ &            &      & $-67\pm30$ & $28\pm 20 $\\
$O_7 =  \frac{3}{N_c^2} S^i T^a G^{ia}$ &  & & & & & $649 \pm 61$ \\ 
\hline
$\chi_{\mathrm{dof}}^2$                                    & $0.43$      & $0.68$ & $0.94$           & $1.04$ & $11.5$ & $0.24$ \\
\hline \hline
\end{tabular}}
\end{table*}

\begin{table*}[h!]
\begin{center}
\caption{Matrix elements of $O_i$ for all states belonging to the 
$[{\bf 70},1^-]$ multiplet.}
\label{Matrix}
\renewcommand{\arraystretch}{2.3} {\scriptsize
\begin{tabular}{lcccccccc}
\hline
\hline
  &\hspace{0cm}  $O_1$ &\hspace{0cm} $O_2$ & \hspace{0cm}$O_3$ &&\hspace{0cm}
  $O_4$ & \hspace{0cm}$O_5$ & \hspace{0cm}$O_6$ & \hspace{0cm}$O_7$\\
  \hline
$^2N_{\frac{1}{2}}$ & $N_c$  & $-\frac{2N_c-3}{3N_c}$ & $\frac{3}{4N_c}$ && $\frac{3}{4N_c}$ & 0 & $\frac{N_c-6}{4N_c}$ &$-\frac{3(N_c-6)}{16N_c^2}$\\
$^4N_{\frac{1}{2}}$ & $N_c$   &  $-\frac{5}{6}$ & $\frac{15}{4N_c}$ && $\frac{3}{4Nc}$ & $\frac{25}{24}$ & $-\frac{5}{8}$ & $\frac{15}{16 N_c}$\\
$^2N_{\frac{3}{2}}$ & $N_c$  & $\frac{2N_c-3}{6N_c}$ &
$\frac{3}{4N_c}$  && $\frac{3}{4N_c}$ & 0 & $-\frac{N_c-6}{8N_c}$ &
$-\frac{3(N_c-6)}{16N_c^2}$\\
$^4N_{\frac{3}{2}}$ & $N_c$  &  $-\frac{1}{3}$ & $\frac{15}{4N_c}$ &&
$\frac{3}{4N_c}$ & $-\frac{5}{6}$ & $-\frac{1}{4}$ & $\frac{15}{16 N_c}$\\
$^4N_{\frac{5}{2}}$ & $N_c$ & $\frac{1}{2}$ & $\frac{15}{4N_c}$ &&
$\frac{3}{4N_c}$ & $\frac{5}{24}$ & $\frac{3}{8}$ & $\frac{15}{16 N_c}$\\
$^2\Delta_{\frac{1}{2}}$ & $N_c$ & $\frac{1}{3}$ & $\frac{3}{4N_c}$ &&
$\frac{15}{4N_c}$ & 0 & $-\frac{5}{4}$ & $\frac{15}{16 N_c}$\\
$^2\Delta_{\frac{3}{2}}$ & $N_c$ & $-\frac{1}{6}$ & $\frac{3}{4N_c}$ &&
$\frac{15}{4N_c}$ & 0 & $\frac{5}{8}$ & $\frac{15}{16 N_c}$\\
$^4N_{\frac{1}{2}}-$ $^2N_{\frac{1}{2}}$ & 0 & 0 & 0 && 0 &
$-\frac{25}{12N_c}\sqrt{\frac{N_c(N_c+3)}{2}}$ &
$-\frac{1}{2N_c}\sqrt{\frac{N_c(N_c+3)}{2}}$ & 0\\
$^4N_{\frac{3}{2}}-$ $^2N_{\frac{3}{2}}$ & 0 & 0 & 0 && 0 &
$\frac{5}{24N_c}\sqrt{5N_c(N_c+3)}$ & $-\frac{1}{4N_c}\sqrt{5N_c(N_c+3)}$ & 0\vspace{0.15cm} \\
\hline \hline
\end{tabular}}
\end{center}
\end{table*}

\begin{table*}[h!]
\begin{center}
\caption{The partial contribution and the total mass (MeV) predicted by the 
$1/N_c$ expansion using Fit 1. 
The last two columns give  the empirically known masses,
name and status.}\label{MASSES1}
\renewcommand{\arraystretch}{2.5}{\scriptsize
\begin{tabular}{lrrrrrrrrrr}\hline \hline
                    &      \multicolumn{6}{c}{Part. contrib. (MeV)}  & \hspace{1.05cm} Total (MeV)   & \hspace{1.05cm}  Exp. (MeV)\hspace{1.05cm}& &\hspace{1.05cm}  Name, status \hspace{.0cm} \\

\cline{2-7}
                    &   \hspace{.35cm}   $c_1O_1$  & \hspace{.35cm}  $c_2O_2$ & \hspace{.35cm}$c_3O_3$ &\hspace{.35cm}  $c_4O_4$ &\hspace{.35cm}  $c_5O_5$ &\hspace{.35cm} $c_6O_6$   &        \\
\hline
$^2N_{\frac{1}{2}}$        & 1444 & 10 & 40 & 42 & 0 & -8  &   $1529\pm 11$  & $1538\pm18$ & & $S_{11}(1535)$****  \\
$^4N_{\frac{1}{2}}$        & 1444 &  26 & 201& 42 & -31& -20 &   $1663\pm 20$  & $1660\pm20$ & & $S_{11}(1650)$**** \\
$^2N_{\frac{3}{2}}$        & 1444 & -5  & 40 & 42 & 0  &  4  &   $1525\pm 8$   & $1523\pm8$  & & $D_{13}(1520)$****\\
$^4N_{\frac{3}{2}}$        & 1444 & 10  & 201& 42 & 25 & -8 &   $1714\pm45$   & $1700\pm50$ & & $D_{13}(1700)$***\\
$^4N_{\frac{5}{2}}$        & 1444 & -16 & 201& 42  & -6 & 12 &   $1677\pm8$    & $1678\pm8$  & & $D_{15}(1675)$****\\
\hline
$^2\Delta_{\frac{1}{2}}$  &  1444 & -10  & 40 & 211 & 0  & -40   & $1645\pm30$  & $1645\pm30$ & & $S_{31}(1620)$**** \\
$^2\Delta_{\frac{3}{2}}$  &  1444 & 5  & 40 & 211 & 0  & 20   & $1720\pm50$  & $1720\pm50$ & & $D_{33}(1700)$**** \\ 
\hline \hline
\end{tabular}}
\end{center}
\end{table*}

\begin{table*}[h!]
\begin{center}
\caption{The partial contribution and the total mass (MeV) predicted by the 
$1/N_c$ expansion using Fit 6. 
The last two columns give  the empirically known masses,
name and status.}\label{MASSES2}
\renewcommand{\arraystretch}{2.5}{\scriptsize
\begin{tabular}{lrrrrrrrrr}\hline \hline
                    &      \multicolumn{5}{c}{Part. contrib. (MeV)}  & \hspace{1.05cm} Total (MeV)   & \hspace{1.05cm}  Exp. (MeV)\hspace{1.05cm}& &\hspace{1.05cm}  Name, status \hspace{.0cm} \\

\cline{2-6}
                    &   \hspace{.35cm}   $c_1O_1$  & \hspace{.35cm}  $c_2O_2$ & \hspace{.35cm}$c_5O_5$ &\hspace{.35cm}  $c_6O_6$ &\hspace{.35cm}  $c_7O_7$  &        \\
\hline
$^2N_{\frac{1}{2}}$        & 1486 & 10  & 0   & -7  & 41 &   $1529\pm 11$  & $1538\pm18$ & & $S_{11}(1535)$****  \\
$^4N_{\frac{1}{2}}$        & 1486 &  25 & -33 & -18 & 203&    $1663\pm 20$  & $1660\pm20$ & & $S_{11}(1650)$**** \\
$^2N_{\frac{3}{2}}$        & 1486 & -5  & 0   & 4   & 41  &     $1525\pm 7$   & $1523\pm8$  & & $D_{13}(1520)$****\\
$^4N_{\frac{3}{2}}$        & 1486 & 10  & 26  & -7  & 203 &    $1718\pm41$   & $1700\pm50$ & & $D_{13}(1700)$***\\
$^4N_{\frac{5}{2}}$        & 1486 & -15 & 7   & 11  & 203 &   $1677\pm8$    & $1678\pm8$  & & $D_{15}(1675)$****\\
\hline
$^2\Delta_{\frac{1}{2}}$  &  1486 & -10 & 0  & -35  & 203     & $1643\pm29$  & $1645\pm30$ & & $S_{31}(1620)$**** \\
$^2\Delta_{\frac{3}{2}}$  &  1486 & 5   & 0  & 18   & 203    & $1711\pm24$  & $1720\pm50$ & & $D_{33}(1700)$**** \\ 
\hline \hline
\end{tabular}}
\end{center}
\end{table*}

\begin{table*}[t]
\renewcommand{\arraystretch}{2.5}
\caption{Matrix elements of operators from the decoupling scheme \cite{CCGL}
corresponding to the $[{\bf 70},1^-]$ multiplet. The columns $\it{asym}$ 
reproduce results obtained with the asymmetric wave function of Ref.
\cite{CCGL} and the columns $\it{sym}$ show results obtained with the 
exact wave function (\ref{EWF}), 
detailed in Appendix A. }
\label{comparison}
\begin{tabular}{c||c|c|c|c|c|c|c|c|} \cline{2-9}
 & \multicolumn{2}{c|}{$\langle s\cdot S_c\rangle$} & \multicolumn{2}{c|}{$\langle S^2_c\rangle$} & \multicolumn{2}{c|}{$\langle t\cdot T_c\rangle$} & \multicolumn{2}{c|}{$\langle T^2_c\rangle$}  \\ \cline{2-9}
& asym & sym & asym & sym & asym & sym & asym & sym\\ \hline
$^2 8$ & $-\frac{1}{2}$ & $-\frac{1}{2}$ & 1 & 1 & $-\frac{1}{2}$ & $-\frac{1}{2}$ & 1 & 1 \\ 
$^4 8$ & $\frac{1}{2}$  & $\frac{1}{2}$  & 2 & 2  & $-1$ & $-\frac{1}{2}$ & 2 & 1\\ 
$^2 10$ &  $-1$ & $-\frac{1}{2}$& 2 & 1 & $\frac{1}{2}$  & $\frac{1}{2}$ & 2 & 2 \\ \hline
\end{tabular}
\end{table*}

In Table   \ref{operators} 
the first nontrivial operator is the spin-orbit operator $O_2$.
In the spirit of the Hartree picture \cite{WITTEN}, generally adopted 
for the description of baryons,  we identify the 
spin-orbit operator with the single-particle operator 
\begin{equation}\label{spinorbit}
\ell \cdot s = \sum^{N_c}_{i=1} \ell(i) \cdot s(i).
\end{equation}
 Accordingly, 
its matrix elements are of order $N^0_c$.
For simplicity  we ignore 
the two-body part of the spin-orbit operator, denoted by
$1/N_c\left(\ell \cdot S_c\right)$ in Ref. \cite{CCGL},
as being of a lower order
(the lower case indicates operators acting on
the excited quark and the subscript $c$ is attached to those
acting on the core).

The operators $O_3$ and  $O_4$ are two-body and linearly independent. 
However, in the decoupling procedure the 
corresponding isospin-isospin operator 
$t^a T^a_c/N_c$ has always been avoided in the numerical analysis 
\cite{CCGL,SGS}.

To be consistent with Ref. \cite{CCGL} we assume that the operators $O_5$
and $O_6$ are dominantly two-body, which means that they carry a
factor $1/N_c$.  Moreover, as $G^{ia}$
sums coherently, it introduces an extra factor $N_c$ and makes the matrix 
elements of $O_5$  and $O_6$ of order $N^0_c$ as well
(what it matters
in the mass operator are the products $c_5 O_5$ and $c_6 O_6$ and it
will turn out that their contribution is small in any case).

We have also included in the fit the following operator 
\begin{equation}
O_7 =  \frac{3}{N_c^2} S^i T^a G^{ia}, 
\end{equation}
an SU(4) invariant built from products of all generators of SU(4),
$S_i$, $T_a$ and $G_{ia}$. In the core plus excited quark procedure
its counterpart was listed in Table I of the second paper of
Ref. [16]  as $O_{16} = g S_c T_c/N^2_c$ but completely 
ignored in the numerical fit, one reason being that the number of operators
in the mass formula was much too large as compared to the data. The operator 
$O_{16}$ is only a part of $O_7$, as it can be easily seen.  As shown below, 
its matrix elements are of 
order $1/N_c$, like those of the pure spin $O_3$ or pure isospin $O_4$ 
operators. Therefore there is no a priori reason to ignore it.  

Naturally, one should also include the operator 
\begin{equation}
O_8 = \frac{1}{N_c} \ell^{(2)} S \cdot S,
\end{equation}  
also of order $1/N_c$.  
However, in our basis
we found a proportionality relation between expectation values
of two different operators 
\begin{equation}
\langle \ell^{(2)ij} S^i  S^j \rangle = 
12 \langle \ell^{(2)ij} G^{ia} G^{ia}\rangle, 
\end{equation}  
for all states belonging to the $[{\bf 70},1^-]$ multiplet.
This implies that we cannot include $O_8$  independently  
in the fit to the experimental spectrum, because its expectation 
values are proportional to those of $O_5$.

The operators $O_5$, $O_6$ and $O_7$ are normalized to allow their 
coefficients $c_i$ to have a natural size \cite{SGS,GSM}. 
The normalization factors follow from the matrix elements 
of $O_i$ presented in Table \ref{Matrix}.
These matrix elements have been calculated for all available states of the 
multiplet $[{\bf 70},1^-]$ starting from the wave function (\ref{WF}) and 
using the isoscalar factors of Table \ref{ISOSC}.
The general analytic expressions of $O_5$, $O_6$ and $O_7$, up to an obvious 
factor, are given in Appendix D.  
For completeness, in Table \ref{Matrix}, 
we also indicate the off-diagonal matrix elements of $O_5$ and $O_6$.

\section{Results}

We have implemented  the matrix elements of Table III
into the mass formula (\ref{massoperator}) and have performed several
distinct fits of the 
theoretical masses to the experiment \cite{Eidelman:2004wy}.
Each of the six fits corresponds to a selection of operators
$O_i$ used in Eq.  (\ref{massoperator}), such as to cover the most
relevant possibilities, in our view.
  
In this way we have obtained sets of values for the dynamical coefficients 
$c_i$ presented in Table \ref{operators}. 
In Tables \ref{MASSES1} and \ref{MASSES2}
we present the masses of the nonstrange resonances
belonging to the $[{\bf 70},1^-]$ multiplet obtained 
from the coefficients resulting from the
Fit 1 and the Fit 6, which correspond to the lowest 
values of $\chi_{\mathrm{dof}}^2$.
 
In Tables \ref{MASSES1} and \ref{MASSES2} we have also indicated
the partial contribution (without error bars) of each term
present in the total mass. These are obtained from the values of $c_i$
of Table  \ref{operators} and the values of $\langle O_i \rangle$ of
Table  \ref{Matrix}.
The Fit 1, containing all operators but $O_7$, is indeed excellent, 
giving $\chi_{\mathrm{dof}}^2 \simeq $ 0.43.
From Table  \ref{operators} one can see that the values of the 
coefficients $c_3$ and $c_4$ are closed to each other, which shows the importance 
of including $O_4$, besides the usual $O_3$.  In addition,
one can see that  $O_3$ is dominant for the $^4N_{J}$ resonances while   
$O_4$ is dominant for the $^2\Delta_J$ resonances,
the contribution being of about 200 MeV in both cases. This brings a new
aspect into the description of excited states studied so far, where  
the dominant term was always the spin-spin term \cite{MS2}, the isospin term
being absent in the numerical analysis. To get a better idea about the 
role of the 
operator $O_4$ we have also made a fit by removing it from the definition
of the mass operator (\ref{massoperator}). The result is shown in 
Table \ref{operators} column Fit 5. The $\chi_{\mathrm{dof}}^2 $ deteriorates
considerably, becoming 11.5 instead of 0.43. This clearly shows that $O_4$ is crucial in the fit.

The coefficient $c_2$ of the spin-orbit term is small and its magnitude and 
sign remains comparable to that of Ref. \cite{Matagne:2006zf} obtained in the
analysis of the $[{\bf 70},\ell^+]$ multiplet. The value of $c_2$ implies 
a small spin-orbit contribution to the total mass, in agreement
with the general pattern observed for the excited states \cite{MS2} and in 
agreement with constituent quark models. 

The error bars of $c_5$ and $c_6$
are comparable to their central values. 
However, the removal of $O_5$ and/or $O_6$ 
from the mass operator does not deteriorate the fit too badly, as shown
in  Table \ref{operators}, Fits 2--4, the  $\chi_{\mathrm{dof}}^2$
becoming at most 1.04. The  contribution of $O_5$ or of
$O_6$ is  comparable to that of the spin-orbit operator. Note that
the structure of $O_6$ is  related to that of the spin-orbit term, which makes
its small contribution entirely plausible. Thus the contribution of all 
operators containing angular momentum is  
small, which may be a dynamic effect.

Table \ref{MASSES2} shows explicitly the role of the operator $O_7$, never
included so far in numerical fits. One can see that this operator 
plays a dominant role in $^4N_J$ and 
 $^2 \Delta_J$, where it contributes with about 200 MeV to the mass,
value comparable to that of   $O_3$ or $O_4$ in the Fit 1. Including    
$O_3$, $O_4$ and $O_7$ together, their contributions remains equally large
but $c_7$ changes sign and $\chi_{\mathrm{dof}}^2$ 
increases to from 0.24 to about 2. 
This suggests that $O_7$ somehow compensates  for the pure spin $S \cdot S$
or pure isospin $T \cdot T$ operators, or in other words,  plays a kind of 
common role with $O_3$ and $O_4$. 
We consider that more theoretical work is needed to better understand
the  algebraic relations between various $O_i$ operators, in particular 
to find new operator identities for mixed symmetric states.

\section{Validity of the approximate wave function} 

In Sec. II it was mentioned that all previous studies of the $[{\bf 70},1^-]$ 
multiplet were performed with the asymmetric wave function (3.4)
of the second paper of Ref. \cite{CCGL}. Here we discuss the validity of 
this approximation by comparing matrix elements of the same operators
calculated both with the exact (symmetric)  and the approximate 
(asymmetric) wave function.

First we consider the  operator $O_2$, common to previous and present
calculations.  
It is a one-body operator, defined by Eq. (\ref{spinorbit}).
Its matrix elements can be written as 
\begin{equation}
\langle \ell \cdot s \rangle =
N_c  \langle \ell(N_c) \cdot s(N_c) \rangle, 
\end{equation}
because the orbital-spin-flavor wave function is symmetric.
Thus it is enough to know the matrix element of a single quark operator, say $N_c$.
Let us illustrate the case $N_c = 5$, 
for which the components of the orbital wave function are given in Table 
\ref{nc=5basisfunctions}.
One can see that only the first basis vector, associated to the 
normal Young tableau,  gives a
nonvanishing contribution to $\langle \ell(N_c) \cdot s(N_c) \rangle$
and this comes only from the term $ssssp$,
because it is the only one where the particle 5 is in a $p$ state.
The generalization of this argument to an arbitrary $N_c$ is obvious
and equally good.
Thus in the case of a single excited quark it is equally well to calculate
$\langle \ell \cdot s \rangle $
with the exact or with the approximate wave function
of Ref. \cite{CCGL}, because the missing terms in the latter function
do not contribute. Note however that
the spin-orbit operator has a negligible contribution to
the mass in all previous and present calculations and in practice it can
be neglected. 

Next we consider a few of the two-body operators of Ref. \cite{CCGL}  $s \cdot S_c$, $S^2_c$, $t \cdot T_c$ and $T^2_c$ 
and restrict the discussion to the case of physical interest,
$N_c$ = 3, which is enough for our purpose.
Appendix A gives the approximate wave functions \cite{CCGL}
and the exact wave functions for the  submultiplets $^2 8$,
$^4 8$ and $^2 10$.  The calculated matrix elements are shown in
Table \ref{comparison}. One can see that for every operator
there is a case where the approximation fails.  
This failure is related to those missing parts of the wave function,
where the core has $I_c \neq S_c$. 
Moreover, the approximate matrix 
elements of the operators $s \cdot S_c$ and $S^2_c$ 
turn out to be isospin dependent and the approximate matrix elements of
the operators $t \cdot T_c$ and $T^2_c$ are spin dependent. 
Using the exact wave function, this anomaly disappears.

For an arbitrary $N_c$ we expect that the exact wave function would generally
give a dependence on $N_c$ for the matrix elements of operators from
previous works, entirely different from that of  Table II and III
of Ref. \cite{CCGL}. 
As a by-product, one can also see
that the operator $T^2_c$, always ignored previously, has matrix elements
comparable to those of $S^2_c$.  This is consistent with our result that
the isospin-isospin becomes as dominant in $\Delta$ as the spin-spin in $N$
resonances.

\section{Conclusions}

In this work we have studied the $[{\bf 70},1^-]$ multiplet in the  
$1/N_c$ expansion by using a simple approach which avoids 
the separation of the system into a core and an excited quark.
This allows us to use an exact wave function
of a system of $N_c$ quarks where both the orbital and spin-flavor
parts are in
the mixed representation $[N_c - 1,1]$. Previously the
wave function was truncated to a single term which implied that 
the symmetry $S_{N_c}$ was broken. 
That asymmetric wave function was associated with describing 
the system as being decoupled into an excited quark and a core
with $N_c - 1$ quarks in the ground state. 
Such a description
 involves an excessively large number of linearly independent 
 operators in the mass formula and the only possible way to
 make a fit was to arbitrarily select some of them. 
 Not surprisingly, the asymmetric wave function fails to reproduce
 the exact values of the matrix elements of some dominant operators  
 in the decoupling scheme itself. 
 
The present approach, based on a wave function with the correct 
permutation symmetry,
sheds an entirely new light on the description 
of the baryon multiplet  $[{\bf 70},1^-]$ in the  $1/N_c$ expansion. 
We have shown that the isospin operator $O_4$ 
is crucial in the fit to the existing data and its contribution to the mass 
of the $\Delta$ resonances  is as important as that 
 of the spin operator $O_3$ for $N$ resonances. 
 Also we found that the operator $O_7$, never included previous fits and
 containing products of all generators 
 of SU(4), Eqs. (5), plays by itself a dominant role in $^4N$ and 
 $^2 \Delta$, states where the spin and isospin are different. 
 By contrast, all operators containing the O(3) generators
 $\ell_i$ bring negligible contributions to the mass.

A comment is in order regarding Refs. \cite{Pirjol:2003ye,Cohen:2003tb} 
where a submultiplet structure (distinct towers of states) has been found,
in the procedure of decoupling the system into a core plus an excited quark.
The present analysis would give similar results. 
The reason is that the existence of three towers of 
states in the $[{\bf 70},1^-]$ multiplet is due to the the presence of three
operators when working up to order $N^0_c$: 
$\openone$ (of order $N_c$) and
$\ell \cdot s$ and $\ell^{(2)} G G/N_c$  
(of order $N^0_c$). 
The meson-baryon scattering analysis of Ref. \cite{Cohen:2003tb} proves
the compatibility between the three towers and three resonance poles in the
scattering amplitude with quantum numbers corresponding to the states
in the $[{\bf 70},1^-]$ multiplet.

  It would be interesting to reconsider the study of higher excited 
  baryons, for example those belonging to $[{\bf 70},\ell^+]$ multiplets, 
  in the spirit of the present approach.
  
  In practical terms, the extension to
  three flavors would involve a considerable amount of
  work on isoscalar factors of SU(6) generators for mixed
  symmetric representations.


\appendix

\section{} 

We consider the particular case of $N_c = 3$ to prove that the wave function
given by Eq. (3.4) of the first paper of Ref. \cite{CCGL} breaks $S_3$ symmetry.

The basis vectors which span the invariant subspace  
of the  mixed symmetric representation correspond to the following
Young tableaux

\begin{equation}
X^{\lambda} \rightarrow \hspace{-0.2cm}\raisebox{-10.5pt}{\mbox{\begin{Young}
1 & 2  \cr
3 \cr
\end{Young}}}\;, X^{\rho} \rightarrow\hspace{-0.2cm} \raisebox{-10.5pt}{\mbox{\begin{Young}
1 & 3  \cr
2 \cr
\end{Young}}}\;,
\end{equation}
where $X$ = $R,S,F$ and $FS$ are the orbital,
spin, flavor and flavor-spin wave functions respectively.

In the spin space one can construct $|S^{\lambda} \rangle$ and  
$|S^{\rho} \rangle$
by first coupling the spin of quarks 1 and 2 to $S_c$ followed by 
the coupling of  $S_c$ to the spin of the third quark. We explicitly have 
\begin{equation}\label{SPINLAM}
|S^{\lambda} \rangle  =  
\sum_{m_1,m_2}
      \left(\begin{array}{cc|c}
	1    &    1/2  &  1/2   \\
	m_1    &    m_2  &  S_3  
      \end{array}\right)    
       \left|S^c=1;m_1\right\rangle\left|\frac{1}{2};m_2\right\rangle,
 \end{equation}
and
\begin{equation}\label{SPINRHO}
|S^{\rho} \rangle  =  
       \left|S^c=0;m_1=0\right\rangle\left|\frac{1}{2};m_2=S_3\right\rangle,
 \end{equation} 
and equivalently in the isospin space
  \begin{equation}\label{ISOLAM}
|F^{\lambda} \rangle  =  
\sum_{\alpha_1,\alpha_2}
      \left(\begin{array}{cc|c}
	1    &    1/2  &  1/2   \\
	\alpha_1    &   \alpha_2  & I_3  
      \end{array}\right)    
       \left|I^c=1;\alpha_1\right\rangle\left|\frac{1}{2};\alpha_2\right\rangle,
 \end{equation}
and
\begin{equation}\label{ISORHO}
|F^{\rho} \rangle  =  
       \left|I^c=0;\alpha_1=0\right\rangle\left|\frac{1}{2};\alpha_2=I_3\right\rangle.
 \end{equation}  
In the standard notation the states ${(FS)}^{\lambda}$ and ${(FS)}^{\rho}$ can be written as
(see \emph{e.g.} \cite{book})
\begin{eqnarray}
 {(FS)}^{\lambda}_{I=3/2;S=1/2}& = &F^{\mathrm{S}}S^\lambda, \label{pro1} \\
{(FS)}^{\rho}_{I=3/2;S=1/2}& = &F^{\mathrm{S}}S^\rho, \label{pro2} \\
{(FS)}^{\lambda}_{I=1/2;S=3/2}& = &F^\lambda S^{\mathrm{S}}, \label{pro5} \\
{(FS)}^{\rho}_{I=1/2;S=3/2}& = &F^\rho S^{\mathrm{S}},\label{pro6} \\
{(FS)}^{\lambda}_{I=1/2;S=1/2}& =& \sqrt{\frac{1}{2}}\left(F^\lambda S^\lambda -F^\rho S^\rho\right),\label{pro3}  \\
 {(FS)}^{\rho}_{I=1/2;S=1/2}& = & -\sqrt{\frac{1}{2}}\left(F^\lambda S^\rho + F^\rho S^\lambda\right).\label{pro4}
\end{eqnarray}
where $F^S$ and $S^S$ denote symmetric states in isospin and spin respectively.
In this notation  the orbital-flavor-spin wave function of a baryon, 
which must be symmetric under $S_3$, is a 
particular case of Eq. (\ref{EWF}) and can be written as
\begin{equation}\label{sym}
|[3] \rangle = \frac{1}{\sqrt{2}} [ R^{\lambda} {(FS)}^{\lambda}
+ R^{\rho} {(FS)}^{\rho}].
\end{equation}

We wish to rewrite the flavor-spin part of the
wave function (3.4) of the first paper of Ref. \cite{CCGL}, denoted by $|II_3;SS_3 \rangle $
in the above notation. 

Let us first consider the case $I  = 3/2, S = 1/2$. One has
\begin{widetext}
\begin{equation}\label{FS1}
|3/2\;I_3; 1/2\; S_3 \rangle  =  
\sum_{m_1,m_2,\alpha_1,\alpha_2}
      \left(\begin{array}{cc|c}
	S_c    &    1/2  &  1/2   \\
	m_1    &    m_2  & S_3  
      \end{array}\right)    
  \left(\begin{array}{cc|c}
	I_c    &    1/2             &  3/2   \\
	\alpha_1    &    \alpha_2   &  I_3  
     \end{array}\right)c^{\mathrm{MS}}_{--}
     \left|S^c=I^c=1;m_1\alpha_1\right\rangle|1/2,m_2 ;1/2,\alpha_2\rangle,
 \end{equation}
\end{widetext}
where  $c^{\mathrm{MS}}_{--}=1$. The spin-flavor states are factorisable
into spin and isospin, so that due to (\ref{SPINLAM})  
this state is identical to
 (\ref{pro1}). For the case $I  = 1/2, S =  3/2$, one has
\begin{widetext}
\begin{equation}\label{FS2}
|1/2\; I_3, 3/2\; S_3 \rangle  =  
\sum_{m_1,m_2,\alpha_1,\alpha_2}
      \left(\begin{array}{cc|c}
	S_c    &    1/2  & 3/2   \\
	m_1    &    m_2   & S_3  
      \end{array}\right)    
  \left(\begin{array}{cc|c}
	I_c    &    1/2   & 1/2   \\
	\alpha_1    &    \alpha_2   & I_3  
     \end{array}\right)c^{\mathrm{MS}}_{++} 
     \left|S^c=I^c=1;m_1\alpha_1\right\rangle|1/2,m_2;1/2;\alpha_2\rangle, 
 \end{equation}
where $c^{\mathrm{MS}}_{++}=1$. Due to (\ref{ISOLAM})
this state is identical to (\ref{pro5}).
\end{widetext}
Next we consider the case $I = 1/2, S = 1/2$, 
\begin{widetext}
\begin{eqnarray}\label{FS3}
|1/2\; I_3, 1/2\; S_3 \rangle &  =  & 
\sum_{m_1,\alpha_1,\eta}
      \left(\begin{array}{cc|c}
	S_c    &   1/2  & 1/2  \\
	m_1    &    m_2   & S_3  
      \end{array}\right)    
  \left(\begin{array}{cc|c}
	I_c    &    1/2   & 1/2   \\
	\alpha_1    &    \alpha_2   & I_3  

      \end{array}\right) c^{\mathrm{MS}}_{0\eta} 
      \left|S^c=I^c=\frac{1}{2}+\frac{\eta}{2};m_1\alpha_1
      \right\rangle|1/2,m_2;1/2,\alpha_2\rangle \nonumber \\
& = & \sqrt{\frac{1}{2}}\left\{\sum_{m_1,\alpha_1}
\left(\begin{array}{cc|c}
	1    &    1/2  & 1/2   \\
	m_1    &    m_2   & S_3  
     \end{array}\right)
\left(\begin{array}{cc|c}
	1    &    1/2  & 1/2   \\
	\alpha_1    &    \alpha_2   & I_3 
 \end{array}\right)\left|S^c=I^c=1; m_1\alpha_1\right\rangle
 |1/2,m_2;1/2,\alpha_2\rangle\right. \nonumber \\
& & \left.\begin{array}{l}
\\ \raisebox{0.29cm}[0cm][0cm]{$
 -  \left|S^c=I^c=0; m_1=\alpha_1=0\right\rangle 
\left|1/2;m_2=\alpha_2=1/2\right\rangle$}
\end{array}\negthinspace\negthinspace\right\},
\end{eqnarray}
\end{widetext}
where we have introduced $c^{\mathrm{MS}}_{0+} = \sqrt{\frac{1}{2}}$ and 
$c^{\mathrm{MS}}_{0-} = - \sqrt{\frac{1}{2}}$
after the second equality sign. 
Due to (\ref{SPINLAM})-(\ref{ISORHO}) this state is 
identical to (\ref{pro3}). This proves 
that in (\ref{FS1}), (\ref{FS2}) and (\ref{FS3})
the second term of Eq. (\ref{sym}) 
is missing. Thus the wave function of Ref. \cite{CCGL}. 
is  truncated.  It contains only one term instead of two   
as required by the $S_3$ symmetry. 
In Sec. VI we show that the missing 
terms  (\ref{pro2}), (\ref{pro6}) and (\ref{pro4}) have a 
considerable contribution to the matrix elements of some operators 
used in the $1/N_c$ expansion mass formula.

\section{}

As an example, in this Appendix we present the orbital basis vectors 
which span the invariant subspace of the representation $[41]$ of $S_5$. 
 
An exact orbital-spin-flavor wave function of 
five fermions (for which the color part is totally antisymmetric)
having the configuration $s^4 p$, \emph{i.e.} a single quark excited 
to the $p$ shell,
has to be built from 4 independent basis vectors, each having a distinct 
Young tableau, both in the orbital 
and spin-flavor spaces.  The basis vectors in the orbital space
are shown in Table \ref{nc=5basisfunctions} \cite{NM}. Note that every term in each state implies
the normal order of particles: $1,2,3,4,5$.  
One can see that the first basis vector, with the 5-th particle in the second
row contains the configuration $ssssp$, \emph{i.e.} it is the only part
of all these basis vectors which has the first four quarks in the
ground state and the 5-th in a $p$ state. One can see that in fact
any quark can be excited to the $p$ shell in a properly
symmetrized state. Thus the wave function
used in previous literature should contain only this $ssssp$ term  
\cite{Goi97,PY1,PY2,CCGL,Pirjol:2003ye,SGS}
if the core was unexcited.
The truncation of the spin-flavor part was 
discussed in Section II.

\begin{table*}[t]
 \caption{Young tableaux and the corresponding basis vectors of the 
 irrep $[41]$  of S$_{5}$ for the configuration $s^4 p$ \cite{book}.}
 \label{nc=5basisfunctions}
 \begin{center}
  \begin{tabular}{l|c}
   \hline \hline \\ \vspace{-0.6cm} &     \\
 \hspace{0.15cm}  Young tableau & \hspace{0.2cm}Young-Yamanouchi basis vectors of [41] \\ \\ \vspace{-0.6cm} &       \\
   \hline \\
\raisebox{-10.5pt}{\mbox{\begin{Young}
1 & 2 & 3 & 4 \cr
5 \cr
\end{Young}}} \hspace{0.5cm} & \hspace{0.2cm}
$\frac{1}{\sqrt{20}} \left(4ssssp-sssps-sspss-spsss-pssss\right)$ \\   \\
\raisebox{-10.5pt}{\mbox{\begin{Young}
1 & 2 & 3 & 5 \cr
4 \cr
\end{Young}}} &
$\frac{1}{\sqrt{12}}\left(3sssps-sspss-spsss-pssss\right)$ \\ \\
\raisebox{-10.5pt}{\mbox{\begin{Young}
1 & 2 & 4 & 5\cr
3 \cr
\end{Young}}} &
$\frac{1}{\sqrt{6}}\left(2sspss-spsss-pssss\right)$ \\ \\
\raisebox{-10.5pt}{\mbox{\begin{Young}
1 & 3 & 4 & 5 \cr
 2 \cr
\end{Young}}} &
$\frac{1}{\sqrt{2}} \left(spsss-pssss\right)$ \\ \\ 
\hline\hline
  \end{tabular}
 \end{center}
\end{table*}


\section{}
The grouping in Table \ref{ISOSC} is
justified by the observation that the isoscalar factors obey the following 
orthogonality relation
\begin{widetext}
\begin{equation}
\sum_{S_1 I_1 S_2 I_2}
 \left(\begin{array}{cc||c}
	[N_c-1,1]    &  [211]   & [N_c-1,1]   \\
	S_1 I_1  &   S_2 I_2   &   S I
      \end{array}\right)_{\rho}
   \left(\begin{array}{cc||c}
	[N_c-1,1]    &  [211]   & [N_c-1,1]   \\
	S_1 I_1  &   S_2 I_2   &   S' I'
      \end{array}\right)_{\rho} = \delta_{S S'} \delta_{I I'},   
 \end{equation}
 \end{widetext}
 which can be easily checked. For example, by taking $S = S'$  and $I = I'$
 one can find that the squares of the first 13 coefficients sum up to one.
 

For completeness also note that the isoscalar factors obey the 
following symmetry property

\begin{widetext}
\begin{equation}
 \left(\begin{array}{cc||c}                                         [N_c-1,1]  &  [211]  &  [N_c-1,1] \\
                           I_1S_1 & I_2S_2 & (S-1)S
                                      \end{array}\right)_{\rho} =
\left(\begin{array}{cc||c}                                         [N_c-1,1]  &  [211]  &  [N_c-1,1] \\
                           S_1I_1 & S_2I_2 & S(S-1)
                                      \end{array}\right)_{\rho}.
\end{equation}
\end{widetext}				      
  				      


\section{}
Here we present the analytic form of the matrix elements of operators proportional to
$O_5$ and $O_6$. They have been obtained following the approach described in Sec. III.
In that notation we have
\vspace{0.5cm}
\begin{widetext}
 \begin{eqnarray}
 \lefteqn{  \langle \ell'S'J'J'_3;I'I_3'|\ell^{(2)ij}G^{ia}G^{ja}|\ell SJJ_3;II_3\rangle = \delta_{J'J} \delta_{J_3J_3'} \delta_{\ell'\ell}\delta_{I'I} \delta_{I_3'I_3}} \nonumber \\ & & 
  \times (-1)^{J+\ell-S}\frac{N_c(3N_c+4)}{16} \sqrt{2S'+1} \sqrt{\frac{5\ell(\ell+1)(2\ell-1)(2\ell+1)(2\ell+3)}{6}}\left\{
  \begin{array}{ccc}
   \ell & \ell & 2 \\
   S & S' & J
  \end{array}\right\}
  \sum_{S''I''}\left\{\begin{array}{ccc}
   1 & 1 & 2 \\
   S & S' & S''
  \end{array}\right\}\nonumber \\ & & \sqrt{\frac{(2S''+1)(2I''+1)}{2I+1}} 
  \times
  \left(\begin{array}{cc||c}
         [N_c-1,1] & [21^2] & [N_c-1,1] \\
	 SI & 11 & S''I'' 
        \end{array}\right)_1 
    \left(\begin{array}{cc||c}
         [N_c-1,1] & [21^2] & [N_c-1,1] \\
	 S''I'' & 11 & S'I 
        \end{array}\right)_1 ,
 \end{eqnarray}

 \begin{eqnarray}
 \lefteqn{ \langle \ell'S'J'J'_3;I'I_3'|\ell^iT^aG^{ia}|\ell SJJ_3;II_3\rangle = \delta_{J'J} \delta_{J_3J_3'} \delta_{\ell'\ell} \delta_{I'I} \delta_{I_3'I_3} (-1)^{J+\ell+S'} \frac{N_c(3N_c+4)}{8} \sqrt{2S'+1}} \nonumber \\& &
 \times \sqrt{\ell(\ell+1)(2\ell+1)}\left\{
  \begin{array}{ccc}
   \ell & \ell & 1 \\
   S' & S & J
  \end{array}\right\}
  \left(\begin{array}{cc||c}
         [N_c-1,1] & [21^2] & [N_c-1,1] \\
	 SI & 11 & S'I 
        \end{array}\right)_1
  \left(\begin{array}{cc||c}
         [N_c-1,1] & [21^2] & [N_c-1,1] \\
	 S'I & 01 & S'I 
        \end{array}\right)_1,
 \end{eqnarray}

\noindent and

\begin{eqnarray}
 \lefteqn{\langle \ell'S'J'J'_3;I'I_3'|S^iT^aG^{ia}|\ell SJJ_3;II_3\rangle = \delta_{J'J} \delta_{J'_3J_3} \delta_{\ell'\ell} \delta_{S'S} \delta_{S_3'S_3}\delta_{I'I} \delta_{I_3'I_3}}\nonumber \\ & & \times \frac{1}{4}\sqrt{N_c(3N_c+4)}\sqrt{I(I+1)}\sqrt{S(S+1)}
  \left(\begin{array}{cc||c}
         [N_c-1,1] & [21^2] & [N_c-1,1] \\
	 SI & 11 & S'I' 
        \end{array}\right)_1.
\end{eqnarray}

\end{widetext}


\vspace{2cm} 
 
{\bf Acknowledgments}
 The work of one of us (N. M.) was supported by the Institut Interuniversitaire des Sciences Nucl\'eaires (Belgium).



\end{document}